\documentclass[aps,prb,twocolumn,10pt,a4paper,footinbib,floatfix]{revtex4}
\usepackage{graphicx,amsmath}
\usepackage{amssymb}
\usepackage{amsthm}
\usepackage{hyperref}
\usepackage{doi}
\allowdisplaybreaks[4]

\begin{document}
\title{Exact results for a $Z_3$ clock-type model and some close relatives}
\author{Iman Mahyaeh}
\author{Eddy Ardonne}
\affiliation{Department of Physics, Stockholm University, SE-106 91 Stockholm, Sweden}
\date{\today}

\begin{abstract}
In this paper, we generalized the Peschel-Emery line of the interacting transverse field Ising model to a model based on three-state clock variables. Along this line, the model has exactly degenerate ground states, which can be written as product states. In addition, we present operators that transform these ground states into each other. Such operators are also presented for the Peschel-Emery case. We numerically show that the generalized model is gapped. Furthermore, we study the spin-S generalization of interacting Ising model and show that along a Peschel-Emery line they also have degenerate ground states. We discuss some examples of excited states that can be obtained exactly for all these models.
\end{abstract}
\maketitle

\section{Introduction}
\label{introduction}
Kitaev's work on Majorana bound states (MBS) \cite{kitaev01} spurred the current interest in zero modes
in general. This resulted in proposals to detect MBSs in nanowires \cite{oreg10,lytchyn10},
resulting in several promising experiments \cite{mourik12,deng12,das12},
trying to observe these zero modes, which if observed, could be used
for (topological) quantum information purposes \cite{kitaev06}.

From a theoretical point of view, one can divide zero modes in two types \cite{af16}.
A zero mode is weak, if it is only associated with a degeneracy of the ground state, while a
strong zero mode implies that the whole spectrum is degenerate (up to corrections that are
exponentially small in the size of the system). Zero modes of non-interacting systems are
strong, as for instance the MBSs of the non-interacting Kitaev chain. Examples of interacting
systems with a strong zero mode are the XYZ chain \cite{paul16} and the {\em chiral} 3-state
Potts model \cite{paul12}. The zero-modes of the later model are interesting, because they
are closely related to parafermionic zero-modes, which are more powerful in comparison to
the MBS, and there are proposals to realize parafermions \cite{mong14,vaezi14,af16}.

In this paper we are interested in interacting systems, that can be fine tuned
such that they have an {\em exact} zero mode for arbitrary system size, i.e., models which have
an exact degeneracy of the ground state. Generic excited states of these models are not degenerate.

Famous examples of models with an exact zero mode are the AKLT \cite{aklt87,aklt88} and Majumdar-Ghosh
(MG) spin chains \cite{mg69a,mg69b}, as well as the interacting transverse field Ising model, along the so-called
Peschel-Emery (PE) line \cite{pe81}. The common denominator of these models is that their ground states
are frustration free. These ground states minimize the energy for each term in the Hamiltonian,
even though these terms in the Hamiltonian do not commute with one another. Obviously, to
achieve this, one has to fine tune the model. This is nevertheless a useful exercise, because
for these fine tuned models, one can often prove some results, such as the existence of
a gap, which is typically impossible for generic Hamiltonians.

We show that the PE-line can be generalized to a
model build from 3-state clock variables, such as the three state Potts model, as 
considered by Peschel and Truong\cite{peschel-truong86}.
Along this line,
the three ground states are exactly degenerate, and can be written as product states (which is not
possible in the AKLT and MG cases). In addition,
we construct edge operators, that permutes these ground states, all along this line. We also construct
such an operator for the PE line, which was not known previously, and present some exact excited
states of these models. We show numerically that the model has a gap.
Finally we introduce a spin-$S$ generalization of the PE-line.

\section{The Peschel-Emery line}
\label{sec:PEline}

The Hamiltonians we consider in this paper are all written as a sum of two-body terms of a $L$-site chain,
\begin{equation}
\label{eq:gen-ham}
H = \sum_{j} h_{j,j+1} \ ,
\end{equation}
where the range of the sum depends on whether we consider an open or closed chain.
For the Ising model in a magnetic field and pair interactions, 
Peschel and Emery \cite{pe81} found that if one parametrizes $h_{j,j+1}(l)$ as follows,
\begin{equation}
\label{hamil EP}
\begin{split}
h_{j,j+1}^{\rm PE} (l) = 
-\sigma^x_j \sigma ^x_{j+1}
+ \frac{h(l)}{2}(\sigma^z_j+\sigma^z_{j+1})\\
+ U(l)\sigma^z_{j}\sigma^z_{j+1} + (U(l) + 1) \ ,
\end{split} 
\end{equation}
the model has two exactly degenerate ground states (with zero energy),
which can be written as product states.
Here, the $\sigma^\alpha$ are the Pauli matrices and
$U(l)=\dfrac{1}{2}[\cosh(l)-1]$, $h(l)= \sinh(l)$ (we note that the sign of $h(l)$ is immaterial)
and $l\geq0$. The model is $\mathbb{Z}_2$ symmetric, with the parity given by
$P= \prod_{j=1}^L \sigma^z_j$. In the open case, the magnetic field of
the boundary spins is half that of the bulk spins.

A direct way to obtain $h_{j,j+1}^{\rm PE}$ was given by Katsura et al. \cite{katsura15}. For the
two site problem, one first demands that the energy of the ground states in the even and odd sectors
are equal, fixing the form of $h(l)$ and $U(l)$. Then one combines the two ground states to write
them as product states. This ensures that the ground states of a chain of arbitrary length
$L$ are frustration free and can be written as product states.
For both the open and periodic chain, they take the form
\begin{equation*}
\vert \psi_{1} (l) \rangle = (\vert\uparrow  \rangle + e^{\frac{l}{2}} \vert\downarrow\rangle)^{\otimes L},
\vert \psi_{2} (l) \rangle = (\vert\uparrow  \rangle - e^{\frac{l}{2}} \vert\downarrow\rangle)^{\otimes L} \ .
\end{equation*}
We note that the energy per bond is $\epsilon(l) = 0$, because of the constant energy shift in Eq.~\eqref{hamil EP}.
These product states do not have definite parity, but orthonormal parity states are constructed as
\begin{align}
\label{partitygs}
\vert E=0 ; \pm \rangle &= \mathcal{N}_{\pm}(l)(\vert \psi_{1}(l) \rangle \pm \vert \psi_{2} (l)\rangle) \ , \\
\mathcal{N}_{\pm}(l) &=  \Big[ 2 (1+ e^l)^L \pm 2 (1-e^l)^L\Big]^{-\frac{1}{2}} \ . 
\end{align}
We label parity eigenstates by both the energy, and the parity eigenvalue.

\subsection{Completely local edge-modes}

The fermionic incarnation of the model Eq.~\eqref{hamil EP}, obtained after performing
a Jordan-Wigner transformation \cite{jw28}, is the Kitaev chain with a nearest-neighbor Hubbard term \cite{katsura15}.
Along the PE-line, this model is in the topological phase \cite{katsura15, alt11}, and has exact zero modes in the open case.
For $U=0$ and arbitrary $h$ the fermionic model is quadratic and can be solved exactly \cite{lsm61,epw70,pfeuty70}.
For $\vert h\vert <1$ the model is topological and hosts two, zero energy, Majorana bound states, localized at the edges
\cite{kitaev01}.
The presence of this zero mode implies that the full spectrum is degenerate up to exponentially
small corrections in the system size.
Generically, upon adding the interaction term, one loses the degeneracy of the full
spectrum \cite{paul17} but as long as one is in the topological phase, the ground state
remains degenerate. The system then has a {\em weak} zero mode, that resides on
the edges of the system, and maps the degenerate ground states into each other\cite{bn16}.

We now construct edge operators, that are completely localized on the edges of the system,
along the full PE-line, but it is insightful to first consider the free fermion point $l=0$.
Using fermion language, such that we associated to Majorana operators $\gamma_{A,j}$
and $\gamma_{B,j}$ to each site $j$, the Majorana edge modes are completely localized
on the first and last sites for $l=0$. In the spin language one of these has a non-local string
operator owing to the Jordan-Wigner transformation, 
\begin{align}
\gamma_{A,1} &= \sigma^x_1
&
\gamma_{B,L} &= -i P \sigma^x_L.
\end{align}
These Majorana operators anti-commute with $P$ and in the ground state space
$\{ \vert E=0; +\rangle , \vert E=0; - \rangle \}$,
they act as $\sigma^x$ and $\sigma^y$ respectively for $l=0$. 

We want to generalize these operators to arbitrary $l$ such that they still permute the parity eigenstates and are normalized (i.e., square to the identity).
The edge operators that satisfy these conditions are 
\begin{align}
\label{Aleft}
A_{\frac{1}{2}}(l)&= e^{-\frac{l}{2}} \sigma^+_1+ e^{\frac{l}{2}} \sigma^-_1, \\
B_{\frac{1}{2}}(l)&=-i P
\left( e^{-\frac{l}{2}} \sigma^+_L + e^{\frac{l}{2}} \sigma^-_L \right),
\label{R2}
\end{align}
where $\sigma^\pm =\frac{1}{2}(\sigma^x \pm i\sigma^y)$. They indeed act on the parity eigenstates as follows,

\begin{align}
\label{act on gs1}
A_{\frac{1}{2}} \vert +\rangle &=\frac{\mathcal{N}_+}{\mathcal{N}_-}\vert -\rangle,~ A_{\frac{1}{2}} \vert -\rangle =\frac{\mathcal{N}_-}{\mathcal{N}_+} \vert + \rangle, \\
\label{act on gs2}
B_{\frac{1}{2}} \vert +\rangle &=i\frac{\mathcal{N}_+}{\mathcal{N}_-}\vert -\rangle,~ B_{\frac{1}{2}} \vert -\rangle =-i\frac{\mathcal{N}_-}{\mathcal{N}_+} \vert + \rangle \,
\end{align}
where $\vert \pm \rangle$ stand for $\vert E=0; \pm \rangle$ and we dropped the dependence on $l$. 

We note that despite the fact that $A_{\frac{1}{2}}(l)^2=B_{\frac{1}{2}}(l)^2=\mathbf{1}$, $\{A_{\frac{1}{2}}(l),B_{\frac{1}{2}}(l)\}=0$ and $\{A_{\frac{1}{2}}(l),P\}=\{B_{\frac{1}{2}}(l),P\}=0$, 
these are not Majorana operators for finite size systems, because
$A^\dagger_{\frac{1}{2}}(l) \neq A_{\frac{1}{2}}(l)$ and $B^\dagger_{\frac{1}{2}}(l) \neq B_{\frac{1}{2}}(l)$ for $l \neq 0$. Since $A^\dagger_{\frac{1}{2}} (l)$ and $B^\dagger_{\frac{1}{2}} (l)$ do not have a simple action on the
ground state space, it does not seem possible to use them to construct Majorana operators
with the desired action on the ground state space for finite system sizes. Despite this, they do constitute an exact zero-mode, all along the PE-line.

However, in the thermodynamic limit we have, 
\begin{equation}
\lim_ {L \rightarrow \infty }\frac{\mathcal{N}_+}{\mathcal{N}_-} =1,
\end{equation}
which means that $A_{\frac{1}{2}}$ and $B_{\frac{1}{2}}$ acts as $\sigma^x$ and $\sigma^y$ respectively in the ground state manifold. Therefore, in this limit, they are Majorana fermions indeed, provided one uses the fermionic incarnation of the model. This also shows, as is well known, that in the fermionic version of the model, the PE-line lies within the topological phase of the model.

We point out that the operators $A_{\frac{1}{2}}$ and $B_{\frac{1}{2}}$, that are defined on
site one and site $L$ in Eqs.~\eqref{Aleft} and \eqref{R2} respectively, could have been defined
on arbitrary sites, because the ground states are product states. However, if one uses the
Jordan-Wigner transformation (see Eq.~\eqref{jordan-wigner} below) to write the model in its fermionic
incarnation, only the operators
$A_{\frac{1}{2}}$ and $B_{\frac{1}{2}}$ of Eqs.~\eqref{Aleft} and \eqref{R2} become Majorana fermions,
that are localized at the left and right edge respectively. The operators in the bulk would have tails
either to the left or to the right.

As it has been pointed out by Alexandradinata et al\cite{bn16}, to study topological order in the ground state manifold weak zero modes are sufficient. These zero modes capture the necessary algebra and act on the ground state manifold as required. Therefore when they are present, one can understand the degeneracy in the ground state manifold and use them to perform the calculation which is needed in the practical setups like T-junctions for braiding.

We should remark that exact Majorana operators can be constructed along the PE-line\cite{katsura15}.
They are exponentially localized at the edges, and take the following form
\begin{align}
\Gamma_{\rm L} &= \frac{1}{\sqrt{\sum_{j=0}^{L-1} q^{2j}}} \sum_{j=1}^{L} q^{(j-1)} \gamma_{A,j} \\
\Gamma_{\rm R} &= \frac{1}{\sqrt{\sum_{j=0}^{L-1} q^{2j}}} \sum_{j=1}^{L} q^{(L-j)} \gamma_{B,j} \ ,
\end{align}
where $q = -\tanh(l/2)$. For completeness, we state the explicit form of the Majorna
operators $\gamma_{A,j}$ and $\gamma_{B,j}$ in terms of the spin operators,
\begin{align}
\gamma_{A,j} &= \bigl( \prod_{k<j} \sigma^z_k \bigr ) \sigma^x_j &
\gamma_{B,j} &= \bigl( \prod_{k<j} \sigma^z_k \bigr ) \sigma^y_j  \ .
\label{jordan-wigner}
\end{align}

\subsection{Exact excited states}

The Majumdar-Ghosh \cite{mg69a,mg69b} and AKLT \cite{aklt87,aklt88} chains,
which have frustration free ground states, also have excited
states that can be obtained exactly for finite system size, see \cite{casp82} and
\cite{arov89,bn17} respectively.
Along the PE-line, one can also obtain exact excited states, in the case with PBC and an even
number of sites. We start with the eigenstates of $h_{j,j+1} (l)$
\begin{align}
\label{eq:two-site Z2 eigenstates}
|g_+\rangle &= |\uparrow\uparrow\rangle + e^{l} |\downarrow\downarrow\rangle &
|g_-\rangle &= e^{l/2} (|\uparrow\downarrow\rangle + |\downarrow\uparrow\rangle) \\
|e_+\rangle &= e^{l} |\uparrow\uparrow\rangle - |\downarrow\downarrow\rangle &
|e_-\rangle &= e^{l/2} (|\uparrow\downarrow\rangle - |\downarrow\uparrow\rangle)
\end{align}
where the ground states $g_\pm$ of both parity sectors have energy $0$, while
$e_-$ and $e_+$ have energy $2$ and $2+U(l)$ respectively.
For simplicity, we dropped the dependence on $l$. 
For a system with an even number of sites, i.e. $L=2N$, the ground states Eq.~\eqref{partitygs}
can be written as
\begin{equation}
\label{eq:pe-line-parity-gs}
\vert E=0 ; \pm \rangle = 2 \mathcal{N}_{\pm} \sum_{i_1\cdots i_N = \pm } g_{i_1} g_{i_2}\dots  g_{i_{N-1}} g_{i_N} \ ,
\end{equation}
where the sum is over all $2^{N-1}$ configurations $i_j = \pm$, with fixed overall parity. Both these
parity ground states have momentum $K=0$, despite the fact that the expression has a two-site
block structure.
Some exact excited states can be obtained by exchanging a ground state block $g$ by an excited
state block $e_\pm$, and summing over all positions for this block. This can be achieved by using
the operators
\begin{align}
O_j^- &= \sigma^z_{2j-1} - \sigma^z_{2j} &
O_j^+ &= \sigma^+_{2j-1} \sigma^+_{2j} - \sigma^-_{2j-1} \sigma^-_{2j}
\end{align}
that act as (focussing on the case with two sites)
\begin{align}
O^- |g_-\rangle &= 2 |e_-> & 
O^- |g_+\rangle &= 0 \\
O^+ |g_+\rangle &= |e_+> & 
O^+ |g_-\rangle &= 0 \ .
\end{align}

Two parity eigenstates with $E = 4$ can be written as
\begin{align}
\label{eq:e4-states}
\vert E = 4, \pm \rangle &= \sum_{j=1}^{N} O^-_j \vert E=0;\pm\rangle \\\nonumber
&= 4 \mathcal{N}_{\pm} \sum_{j=1}^{N}
\sum_{i_1\cdots i_N = \pm } g_{i_1}\cdots g_{i_{j-1}}e_{-} g_{i_{j+1}} \cdots  g_{i_N} \ ,
\end{align}
where $i_j = -$ is fixed in the second sum. These states automatically have momentum $K=\pi$. 
Exchanging the block $e_-$ by $e_+$ gives two excited states with energy $E = 4 + 4U(l)$. One
starts with
\begin{align*}
| \Psi,\pm \rangle &=  \sum_{j=1}^{N} O^+_j \vert E=0 ; \pm \rangle \\
&=2 \mathcal{N}_{\pm}\sum_{i_1\cdots i_N = \pm } g_{i_1}\cdots g_{i_{j-1}}e_{+} g_{i_{j+1}} \cdots  g_{i_N} \ ,
\end{align*}
and constructs $K=\pi$ states as follows
\begin{equation}
\vert E = 4 + 4 U(l), \pm \rangle =
| \Psi,\pm \rangle - T | \Psi,\pm \rangle \ ,
\end{equation}
where $T$ translates the system by one site. Finally, by introducing both one $e_-$ block and one
$e_+$ block results in the states $| \Psi',\pm \rangle$,
\begin{equation*}
| \Psi',\pm \rangle =  \sum_{j=1}^{N} O^+_j \vert E=4 ; \pm \rangle \ .
\end{equation*}
From these, one obtains two $K=0$ states with energy $E = 8 + 4U(l)$,
\begin{equation}
\vert E = 8 + 4 U(l), \pm \rangle =
| \Psi',\pm \rangle + T | \Psi',\pm \rangle \ .
\end{equation}
In App.~\ref{app:excited-state}, we prove that the states $\vert E=4;\pm\rangle$ are indeed
exact excited states of the Hamiltonian. The proof for the other states works in a similar way.

\section{The 3-state clock model}
\label{3state-clock-case}
The construction of the PE-line can be generalized to 3-state clock or Potts
type models\cite{peschel-truong86}.
The Hamiltonian of the 3-state clock model, which is a generalization of the transverse field Ising model, is
\begin{equation}
\label{clock hamil}
H= -\sum_{j=1}^{L-1}(X_j^{\dagger}X_{j+1} + {\rm h.c.}) - \sum_{j=1}^L (f Z_j^\dagger + {\rm h.c.}) \ .
\end{equation}
To each site, one associates a three-dimensional Hilbert space, $| n \rangle$ with $n=0,1,2$
taken modulo 3.
The clock operators $Z$ and $X$ act as  $Z \vert n\rangle = \omega ^n \vert n \rangle$ with
$\omega = \exp (i \frac{2\pi}{3})$ and $X \vert n \rangle = \vert n-1\rangle$.
These operators satisfy $X^3=Z^3=\mathbf{1}$, $X^2=X^\dagger$, $Z^2=Z^\dagger$ and
$XZ=\omega ZX$.
Although this model is not solvable in general, it is known that for $|f|<1$ this model has three
degenerate ground states (a weak zero mode\cite{paul12}), while for $|f|>1$ it shows a paramagnetic behaviour; the behaviour of the
critical point at $f=1$ is described by the $Z_3$ parafermion CFT, see Ref.~\onlinecite{dotsenko84}.

The clock model Hamiltonian commutes with the parity operator which is now defined as $P=\prod_{j=1}^L Z$, hence Hamiltonian is $\mathbb{Z}_3$ symmetric. Therefore states can be labeled with their parity eigenvalue, $P=\omega^Q$, in which $Q$ could be $0$, $1$ or $2$ since $P^3=\mathbf{1}$. The phase diagram of this model and in particular its chiral generalization\cite{zh15} was recently investigated\cite{paul12,jf14}; in
particular, the presence and stability of parafermionic zero modes was studied. There is consensus that the chiral Potts model hosts a strong
Parafermionic zero mode at $\theta = \pi/6$, but the nature of the zero mode at generic angles is under debate\cite{jf14,Joost17}.

Apart from the integrable points of the model \cite{zamolodchikov85}, the clock model has not been solved.
Recently, Iemini et al \cite{ma17} found a generalization of the model for which the ground state
is exactly three-fold degenerate along a specific line;
moreover, these ground states have a matrix-product form which becomes simple in terms of Fock parafermions \cite{ort14}.
Here, we consider a generalization of the Potts model with fine-tuned couplings, such that
the ground states can be written as a product state, in direct analogy with the PE line for the spin-$1/2$ case. 

\subsection{Construction of the generalized Potts model}

We use the method\cite{katsura15} that we outlined in the previous section.
One first needs to establish which terms to add to the Hamiltonian 
Eq.~\eqref{clock hamil}, in analogy to the Hubbard-U term present in Eq.~\eqref{hamil EP}.
It turns out that one needs both the terms
$Z_{j} Z_{j+1}$ and $Z_{j} Z_{j+1}^\dagger$. With these terms, we consider the following two-site
Hamiltonian in Eq.~\eqref{eq:gen-ham},
\begin{align}
\label{Z3 line}
h_{j,j+1}^{Z_3} (r)&= \bigl[-X^\dagger_j X_{j+1} - f(r)(Z_j + Z_{j+1})\\
&-g_1(r) Z_j Z_{j+1}-g_2(r)Z_j Z^\dagger_{j+1}+ {\rm h.c} \bigr]
+ \epsilon(r) \ ,
\nonumber
\end{align}
where $\epsilon(r) = 2(1+r+r^2)^2/(9r^2)$.
We find that the the following parameters are required to construct a PE-line,
\begin{align}
f(r) &= (1+2r)(1-r^3)/(9r^2)\\
g_1 (r) &= -2(1-r)^2(1+r+r^2)/(9r^2)\\
g_2 (r) &= (1-r)^2(1-2r-2r^2)/(9r^2) \ ,
\end{align}
where $r >0$ and $r=1$ corresponds to the non-interacting model, see also\cite{katsura-potts}.
Note that as for the PE-line, the `magnetic field' term is half as strong on the boundary sites in
comparison to the bulk sites. This model has three exactly degenerate ground states with zero
energy, the latter due to the explicit energy shift $\epsilon(r)$.
These ground states can, by construction,
be written as product states that take the form
\begin{align}
\vert G_0(r) \rangle &= \left( \vert 0 \rangle+ r \vert 1 \rangle + r \vert 2 \rangle \right)^{\otimes L} \\
\vert G_1(r) \rangle &= \left( \vert 0 \rangle+ r \omega \vert 1 \rangle + r \bar{\omega}\vert 2 \rangle \right)^{\otimes L} \\
\vert G_2(r) \rangle &= \left( \vert 0 \rangle+ r \bar{\omega} \vert 1 \rangle + r \omega \vert 2 \rangle \right)^{\otimes L} \ .
\end{align}
These product states can be combined to form orthonormal parity eigenstates,
\begin{align}
\label{z3paritygs}
\vert E=0; 1 \rangle &= \mathcal{N}_1(\vert G_0(r) \rangle+ \vert G_1(r)\rangle + \vert G_2(r)\rangle ) \nonumber\\
\vert E=0; \omega \rangle &= \mathcal{N}_{\omega}(\vert G_0(r)\rangle+ \bar{\omega}\vert G_1(r)\rangle + \omega \vert G_2(r)\rangle ) \\
\vert E=0; \bar\omega \rangle &=\mathcal{N}_{\bar{\omega}} ( \vert G_0(r) \rangle+\omega \vert G_1(r)\rangle + \bar{\omega}\vert G_2(r)\rangle ) \ ,
\end{align}
where 
\begin{align}
\mathcal{N}_1&=\Big[ 3 (1+ 2r^2)^L + 6 (1-r^2)^L\Big]^{-\frac{1}{2}}, \\
\mathcal{N}_{\omega,\bar{\omega}}&=\Big[ 3 (1+ 2r^2)^L - 3 (1-r^2)^L\Big]^{-\frac{1}{2}}.
\end{align}
These states are labeled by their energy and their `parity' eigenvalue of $P$.

\subsection{Completely local edge modes}
As was the case for the PE-line of the spin-$1/2$ model, one can explicitly construct edge operators for the open chain.
For $r=1$, the couplings $f,g_1,g_2$ are zero and we are left with
$h_{j,j+1}^{Z_3} (1) = -X_jX^\dagger_{j+1}+{\rm h.c.}$.
To find the zero-mode operators in this limit, one uses the Fradkin-Kadanoff transformation\cite{fk70}
to transform the clock degrees of freedom to parafermions $\eta_{A,j}$ and $\eta_{B,j}$,
\begin{align}
\label{fradkin-kadanoff}
\eta_{A,j} &= \bigl( \prod_{k<j} Z_k \bigr) X_j  
&
\eta_{B,j} &= \omega \bigl( \prod_{k<j} Z_k \bigr) X_j Z_j \ .
\end{align}
These operators satisfy
\begin{align}
\eta_{A,j}^3 &= \eta_{B,j}^3 = \mathbf{1} \\
\eta_{x,j}^2 &= \eta_{x,j}^\dagger \\
\eta_{x,j} \eta_{x',j'} &= \omega^{{\rm sgn}(j'-j)} \eta_{x',j'} \eta_{x,j} & \text{if $j\neq j'$} \\
\eta_{A,j} \eta_{B,j} &= \omega \eta_{B,j}\eta_{A_j} \ ,
\end{align}
where $x,x'$ are $A$ or $B$. 
One finds that the Hamiltonian does not depend on two of the parafermions \cite{paul12}, namely
\begin{align}
\eta_{A,1} &= X_1 & \eta_{B,L} &= \bar\omega P X_L \ .
\end{align}
These operators obey the parafermion algebra, $\eta_{A,1}^3=\eta_{B,L}^3=\mathbf{1}$ and
$\eta_{A,1}\eta_{B,L}=\omega \eta_{B,L} \eta_{A,1}$.
To find edge modes for arbitrary $r$, we first note that
$\eta _{A,1}$ and $\eta_{B,L}$ act on the ground state space
$\{ |E=0;1\rangle, |E=0;\omega\rangle, |E=0;\bar{\omega}\rangle \}$ (with $r=1$) as
$X$ and $\bar{\omega}ZX$.
To generalize these operators to arbitrary $r$, it is useful to consider the generalization of the
ladder operators for $SU(2)$ spins, namely
\begin{align}
&\Sigma^0=\frac{X}{3}\left( \mathbf{1} + Z + Z^\dagger \right) \\
&\Sigma^1=\frac{X}{3}\left( \mathbf{1} + \bar{\omega}  Z + \omega Z^\dagger \right)   \\
&\Sigma^2 =\frac{X}{3}\left( \mathbf{1} + \omega  Z + \bar{\omega}Z^\dagger \right) \ .
\end{align}
One checks that
$\Sigma^0 \vert 0\rangle = \vert 2 \rangle$,
$\Sigma^1 \vert 1\rangle = \vert 0 \rangle$ and
$\Sigma^2 \vert 2\rangle = \vert 1 \rangle$ while all the other matrix elements are zero.

The edge operators that act in the same way as
$\eta_{A,1}$ and $\eta_{B,L}$
for arbitrary $r$ can be written in terms of the $\Sigma^\alpha$'s as
 \begin{align}
\label{pfopa}
&A_{Z_3}(r) = \frac{1}{r} \Sigma^1_1 +\Sigma^2_1 + r \Sigma^0_1, \\
\label{pfopb}
&B_{Z_3}(r) = \bar\omega P \left(\frac{1}{r} \Sigma^1_L + \Sigma^2_L + r \Sigma^0_L \right) \ .
\end{align}
One can check that,
\begin{align}
A_{Z_3} |1\rangle &= \frac{\mathcal{N}_1}{\mathcal{N}_{\bar{\omega}}} |\bar{\omega}\rangle,~A_{Z_3} |\omega \rangle = \frac{\mathcal{N}_{\omega}}{\mathcal{N}_{1}} |1\rangle,~A_{Z_3} |\bar{\omega}\rangle = | \omega \rangle \ , \\
B_{Z_3} |1\rangle &= \omega \frac{\mathcal{N}_1}{\mathcal{N}_{\bar{\omega}}} |\bar{\omega}\rangle,~B_{Z_3} |\omega \rangle =\bar{\omega} \frac{\mathcal{N}_{\omega}}{\mathcal{N}_{1}} |1\rangle,~B_{Z_3} |\bar{\omega}\rangle = | \omega \rangle \ ,
\end{align}
where $|1,\omega,\bar{\omega} \rangle$ stand for $|E=0;1,\omega,\bar{\omega} \rangle$. 
Although these operators obey the relations $(A_{Z_3})^3=(B_{Z_3})^3=\mathbf{1}$ and
$A_{Z_3}B_{Z_3} = \omega B_{Z_3}A_{Z_3}$, they are not parafermions, because 
for instance $A_{Z_3}^\dagger \neq (A_{Z_3})^2$, and likewise for $B_{Z_3}$.
The situation we encounter here is analogous to the spin-$1/2$ PE-line. If one tries to construct
completely local parafermion operators, one finds that one of the necessary relations is not satisfied.
Despite that, the operators $A_{Z_3}(r)$ and $B_{Z_3} (r)$ are exact zero modes.
However, it is worthwhile to mention that as in the $Z_2$ case, in the thermodynamic limit the ratio
$\frac{\mathcal{N}_{\omega}}{\mathcal{N}_{1}}$ approaches $1$ and we obtain weak parafermionic
zero modes. 

The operators $A_{Z_3}(r)$ and $B_{Z_3} (r)$, Eqs.~\eqref{pfopa} and \eqref{pfopb} are
defined on the first and last site, respectively.
As was the case for the spin-1/2 EP-line, these operators could have been defined on any site,
without changing the way they permute the different ground states. However, if one uses the
Fradkin-Kadanoff transformation\cite{fk70}, Eq.~\eqref{fradkin-kadanoff}, only the operators
Eqs.~\eqref{pfopa} and \eqref{pfopb} become local parafermion operators. In Sec.~\ref{numerics}
below, we numerically show that the model Eq.~\eqref{Z3 line} has a gap between the three-fold
degenerate ground states and the excited states. Together this implies that that model,
in its parafermionic representation, lies within a topological phase for finite values of the parameter $r$.

In the spin-$1/2$
case, it was possible to construct exponentially localized Majorana operators, that do satisfy the
correct algebra for arbitrary finite system size. It is tempting to try to do the same thing for the
current $Z_3$ case. It turns out that this is hard. Even constructing parafermion operators for a
system with only two sites is much harder than
it looks at first sight. In App.~\ref{app:two-site-parafermion-operator}, we construct the most
general, two-site parafermion operator, that satisfies all the required properties. Given the complexity
of the two-site problem, we do not discuss longer chains.

\subsection{Exact excited states}

As was the case for the spin-$1/2$ PE-line, one can construct exact excited states in case of a
system with an even number of sites $L = 2N$ with periodic boundary conditions.
We write the ground state and two excited states of $h^{Z_3}_{j,j+1}$ explicitly, because they are
the building blocks of our construction,
\begin{align}
\vert g_1\rangle &= \vert 00 \rangle + r^2 \vert 12 \rangle + r^2 \vert 21\rangle \ , \\
\vert g_\omega\rangle &= r^2\vert 22 \rangle + r \vert 01 \rangle + r \vert 10\rangle \ , \\
\vert g_{\bar{\omega}}\rangle &= r^2\vert 11 \rangle + r \vert 02 \rangle + r \vert 20\rangle \ , \\
\vert e_\omega\rangle &= 3 r ( \vert 01 \rangle - \vert 10 \rangle ) \ , \\
\vert e_{\bar{\omega}}\rangle &= 3 r ( \vert 02 \rangle - \vert 20 \rangle ) \ ,
\end{align}
where $g_{1,\omega,\bar{\omega}}$ have energy $0$ and
$e_{\omega,\bar{\omega}}$ have energy $2+r$. The excited states are obtained
by acting with the operator $O = Z_1 - Z_2 + h.c.$ on the ground states, namely
\begin{align}
O \vert g_1 \rangle &= 0 & 
O \vert g_\omega \rangle &= \vert e_\omega \rangle & 
O \vert g_{\bar{\omega}} \rangle &= \vert e_{\bar{\omega}} \rangle \ . 
\end{align}

We can rewrite the three ground states in terms of these blocks,
\begin{equation}
\label{eq:z3-parity-gs}
\vert E = 0;1,\omega,\bar{\omega}\rangle =3 \mathcal{N}_{1,\omega,\bar{\omega}}
 \sum_{i_1\cdots i_N = 1,\omega,\bar{\omega} } g_{i_1} g_{i_2}\dots  g_{i_{N-1}} g_{i_N} \ ,
\end{equation}
where the sum is over all $3^{N-1}$ configurations with $i_j = 1,\omega,\bar{\omega}$, and fixed
overall `parity'. There are three exact excited state with energy $\Delta E=2(2+r)$ and momentum
$K=\pi$ along, which can be constructed by acting with the operator
$O_{\rm tot} = \sum_{j=1}^{N} Z_{2j-1} - Z_{2j} + {\rm h.c.}$, 
\begin{align}
&\vert E = 2(r+2); \omega,\bar{\omega}\rangle = \nonumber \\
& \sum_{j=1}^{N} (Z_{2j-1} - Z_{2j}+h.c.)
\vert E=0;1,\omega,\bar{\omega}\rangle \ .
\end{align}
Effectively, the operator replaces one of `$g_i$-blocks' by an `$e_i$-block'
with the same parity $\omega$ or $\bar\omega$, and summing over the possible positions of these
blocks.

\subsection{Numerical results}
\label{numerics}
In this section we present our numerical study of the model, in particular we study the energy gap using  
DMRG\cite{white92,scholl05}, making use of the ALPS libraries\cite{alps07,alps11,alps14}. From this
study, we conclude that the the $Z_3$ model, Eq.~\eqref{Z3 line} is gapped, in analogy to the $Z_2$ case. 

\begin{figure}[t]
	\includegraphics[width=\columnwidth]{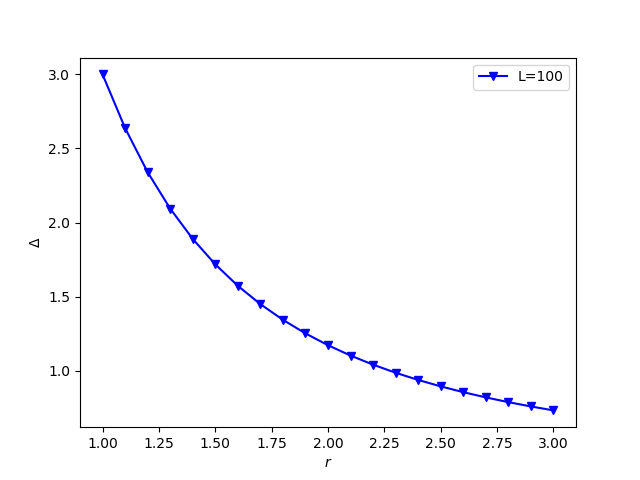}
	\caption{ The bulk gap of the model in Eq.~\eqref{Z3 line} as a function of $r$. We performed DMRG for $L=100$. } 
	\label{fig:gap}
\end{figure}

Since the first three states are degenerate with zero energy, we need to determine the energy of the lowest four eigenstates. Even though this is quite demanding, we were able to do so using ALPS. We performed DMRG calculations to find the gap of the $Z_3$ model, Eq.~\eqref{Z3 line}, for $L=100$ sites with open (free) boundary conditions. We keep up to $\chi=100$ states in the Schmidt decomposition provided their Schmidt eigenvalues are all larger than $10^{-10}$ and we perform three sweeps. To check convergence, we also considered $\chi=200$ and found that the energies were within the current numerical errors.  Based on our numerical results, the first three states have energy of the order $10^{-10}$, which shows that the energy for these (exactly) zero energy states is well converged.
We obtained the energy gap $\Delta$, i.e. the gap to the fourth eigenstate, with an error of the order of $10^{-4}$. The finite size gap for $L=100$ is presented in Fig.~\ref{fig:gap}, for $1.00 \leq r \leq 3.00$. 

To establish the existence of a gap in the thermodynamic limit, we study the size dependence of the gap.
The exact solution for the (non-interacting) transverse field Ising model shows that the finite size gap
converges to its thermodynamic value as $L^{-2}$ in the ordered phase. We checked that for the PE line the
gap saturates to its thermodynamic value as $L^{-\alpha}$ where $\alpha$ is very close to $2$. 

\begin{figure}[t]
	\includegraphics[width=\columnwidth]{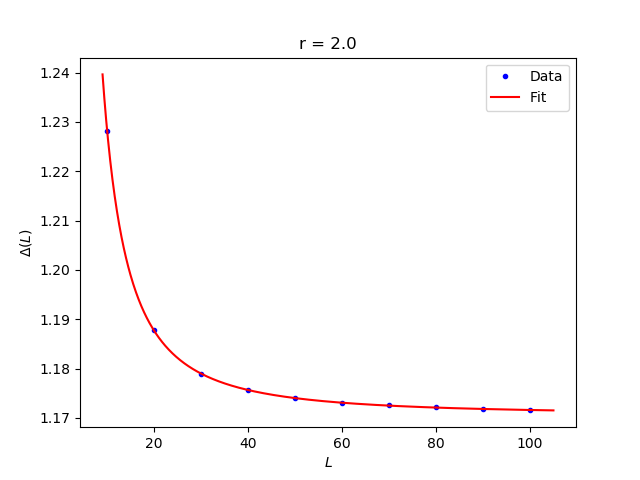}
	\caption{ The bulk gap of the model in Eq.~\eqref{Z3 line} as a function of $r$. We performed DMRG for $L=100$. We plot data and the curve $\Delta(L)= 3.236 L^{-1.751} + 1.1706$. The error for each gap point is $10^{-4}$.} 
	\label{fig:scaling-gap}
\end{figure}

We numerically determined the gap along the $Z_3$ line for different system sizes up to $L=100$. We fitted a power-law function, $\Delta(L)=a L^{-b} + \Delta_{\infty}$ (in analogy with the $Z_2$ case). The data and the fitted curve for $r=2.0$ are presented in Fig.~\ref{fig:scaling-gap}. The gap decays as $L^{-1.75}$ to its thermodynamic value $\Delta_{\infty} = 1.171$.
Recent results on the gap of frustration-free models show that
if the gap decays to a finite value faster than $L^{-3/2}$, the model is gapped \cite{lemm18}, including our model Eq.~\eqref{Z3 line}.

As we mentioned above, the error in the energy is of the order $10^{-4}$ in our calculation. The difference between the gap for $L=90$ and $L=100$ is $2 \times 10^{-4}$ which shows that the energy has basically converged to its final value within our precision. We also checked that the gap converges to a finite non-zero value with the same behaviour and $b \approx 1.75$ in the range $1.00 < r < 3.00$.

We numerically found that the gap decreases as $r$ increases, in analogy to the PE line. As it was pointed out in the previous studies\cite{ds03}, in the large $l$ limit of the PE line, where the $U$ and $h$ couplings dominate the Hamiltonian, the PE-line reaches a multicritical point. At this point the ground state degeneracy grows exponentially with the system size. To see this, following Ref.\cite{ds03}, we rewrite $h^{PE}_{j,j+1}(l)$ for large $l$, 
\begin{equation}
\lim_{l \rightarrow \infty} h^{PE}_{j,j+1}(l) = \frac{e^l}{4} \left( \sigma^z_j + \mathbf{1}\right)\left( \sigma^z_{j+1} + \mathbf{1}\right) \ . 
\end{equation} 
For this Hamiltonian any state which does not have two adjacent spins in the $+z$ direction is a ground state, explaining the exponential degeneracy of the ground state with system size. 

The same thing happens along the $Z_3$ line. In the large $r$ limit we can rewrite the Hamiltonian as
\begin{align}
\lim_{r \rightarrow \infty} h_{j,j+1}^{Z_3}(r) & = \nonumber  \\  
& \frac{2}{9} r^2 \left( \mathbf{1} + Z_j + Z^\dagger_j \right) \left( \mathbf{1} + Z_{j+1} + Z^\dagger_{j+1} \right) \ .
\end{align}
Similar to the PE line in this limit any state which does not have two adjacent "spins" in the $n=0$ state, is a ground state. Therefore we conclude that our model has a multicritical point for $r\rightarrow\infty$.

\section{Spin-$S$ Peschel-Emery line}
\label{spin-s-case}
We study the spin-$S$ generalization of the PE-line which has been investigated
previously \cite{kur82,sen91,muller85,ds03}.
Here we present the exact ground states, which again are product states, as well as 
exact, local edge modes and two exact excited states. 
The Hamiltonian for this model is
\begin{equation}
\label{hamil S}
\begin{split}
h^{\rm S-PE}_{j,j+1} = 
- S^x_j S^x_{j+1} + \frac{h(l)}{2}S \left( S^z_j + S^z_{j+1}\right) \\
+ U(l) S^z_j S^z_{j+1} + S^2 (U(l) + 1)\ ,
\end{split}
\end{equation}
in which $S^\alpha$ are spin-$S$ operators of $SU(2)$.
The parameters
$U(l)=\dfrac{1}{2}[\cosh(l)-1]$ and $h(l)= \sinh(l)$,
are the same as the PE-line couplings in Eq.~\eqref{hamil EP}. We note that
in the $S=1/2$ case, the Hamiltonian Eq.~\eqref{hamil S} is
$\frac{1}{4}$ times $h^{\rm PE}_{j,j+1}$ (see Eq.~\eqref{hamil EP}),
which is written in terms of Pauli operators instead of spin-$1/2$ operators.

The Hamiltonian Eq.~\eqref{hamil S} commutes with the `parity' of the magnetization,
$P_M=\prod_{j=1}^L e^{i \pi (S - S^z_j)}$, because the operators $S^x_{j} S^x_{j+1}$
either change the magnetization by two units, or leave it unchanged.

The model has two exactly degenerate ground states for arbitrary $l$, which can be written as
product states, similar to the $Z_2$ and $Z_3$-clock model cases.
These two ground states are
\begin{align}
\vert \psi_{S,1}(l) \rangle &= \left( e^{\alpha S^-} \vert S \rangle_z \right)^{\otimes  L}
&
\vert \psi_{S,2}(l) \rangle &= \left( e^{-\alpha S^-} \vert S \rangle_z \right)^{\otimes  L} 
\end{align}
where $\alpha =\exp (\frac{l}{2})$ and $\vert S \rangle_z$ is the $S^z = S$ eigenstate,
i.e. $S^z \vert S \rangle_z = S \vert S \rangle_z$. The states
$\vert \psi_{S,1}(l) \rangle$ and $\vert \psi_{S,2}(l) \rangle$ are not parity eigenstates,
but these can be constructed as
\begin{equation}
| E = 0 ; \pm \rangle = (\vert \psi_{S,1} (l) \rangle \pm \vert \psi_{S,2} (l) \rangle )/2 \ .
\end{equation}
As in the previous cases, these states are exact ground states for both the open and
periodic chains, with the energy per bond given by $\epsilon_S(l)= 0$.

Following the $\mathbb{Z}_2$ case we can define local edge operators,
\begin{align}
A_S(l) &= \frac{1}{2S} (\frac{1}{\alpha} S^+_1 + \alpha S^-_1) \\ 
B_S(l) &= - \frac{i}{2S} P_M (\frac{1}{\alpha} S^+_L + \alpha S^-_L) \ . 
\end{align}
For $S = 1/2$, these operators reduce to $A_{\frac{1}{2}}(l)$ and $B_{\frac{1}{2}}(l)$ in Eq.~\eqref{R2}.
They act like $S^z$ and $-S^y$ on the ground states $\{ \vert \psi_{S,1} (l) \rangle,\vert \psi_{S,2} (l) \rangle \}$.

In the case of periodic boundary conditions, it is possible to write exact excited states of the model
Eq.~\eqref{hamil S}. These excited states are constructed from the 
ground states of the model with two sites as before. The ground states $|g_{\pm}\rangle$ with
parities $P_M = \pm 1$ of the two site model are obtained by acting on $| S, S\rangle$ as
\begin{equation}
|g_\pm\rangle = \left[ e^{\alpha (S^-_1+S^-_2)} \pm e^{-\alpha (S^-_1+S^-_2)}\right]  | S, S\rangle \ .
\end{equation}
There are two parity eigenstates $| e_\pm \rangle$ with energy $E = S$, which can be
obtained from the ground states
\begin{equation}
| e_\pm \rangle = (S^z_1 - S^z_2) |g_\pm\rangle \ .
\end{equation}
We note that we assumed that $S>1/2$ here, because for $S=1/2$, we have
$(S^z_1 - S^z_2) |g_+ \rangle = 0$, in agreement with results for the $Z_2$ case discussed above.

To find the two exact excited states of the system with length $L = 2N$, we first rewrite the
ground states of the $L=2N$ site chain in terms of the $g_\pm$, similar to
Eq.~\eqref{eq:pe-line-parity-gs}
\begin{equation}
\label{eq:spin-s-pe-line-parity-gs}
\vert E=0; \pm \rangle = \sum_{i_1\cdots i_N = \pm } g_{i_1} g_{i_2}\dots  g_{i_{N-1}} g_{i_N} \ ,
\end{equation}
where the sum is again over all $2^{N-1}$ configurations $i_j = \pm$, with fixed total parity of
the magnetization.
These states
are ground states for both the open and periodic cases, with momentum $K=0$.
From these $K=0$ states, one obtains $K=\pi$, parity eigenstates with energy $E = 2S$, by
replacing the one of `$g_i$-blocks' by an `$e_i$-block' with the same parity, and summing over the
position, again in analogy with the spin-$1/2$ case,
\begin{equation}
\vert E = 2 S; \pm \rangle =
\sum_{j=1}^{N} (S_{2j-1}^z - S_{2j}^z)
\vert E = 0; \pm \rangle \ .
\end{equation}

\section{Discussion}
\label{discussion}

We considered one-dimensional models for which the ground states and a few excited
states can be obtained analytically. These models are inspired by the Peschel-Emery line\cite{pe81},
of the interacting transverse field Ising model (or, in its fermionic incarnation, Kitaev's
Majorana chain in the presence of a Hubbard interaction). In particular, we constructed
a direct analog of the PE-line, starting from the 3-state Clock/Potts model, by introducing two types
of additional interaction terms.

For the resulting one-parameter family of
models, the three-fold degenerate ground states can be written in product form. In addition, we
found a triple of excited states that can be obtained analytically. More importantly, we constructed
completely local operators, that permute the parity ground states. These operators almost satisfy
the parafermion relations, the only requirement missing is that they are not unitary.
Although we believe it should be possible to construct (exponentially) localized parafermion
operators, we only succeeded in constructing these for the two-site problem, where they already
are quite complicated.

The model studied in this paper behaves in close analogy to the
model considered recently by Iemini et al.\cite{ma17}. It would be interesting to see if both
models can be obtained from a more general model. For instance, it is  
interesting to note\cite{mora-private}
that the construction of the local operators that permute the ground states
can be extended to the model of Ref.~\onlinecite{ma17}.

In addition to the results for the 3-state clock-type models, we also considered an arbitrary
spin-$S$ version of the Peschel-Emery line.

There has been a lot of interest in Clock/Potts type models recently, both the chiral as well
as the non-chiral versions. It was only rather recently that the phase-diagram of the chiral
3-state Potts model has been investigated in detail\cite{zh15}. The additional `interaction' terms that we
needed to consider, namely $Z_j Z_{j+1} + {\rm h.c.}$ and $Z_j Z^\dagger_{j+1} + {\rm h.c.}$
have not attracted much attention yet, but they were considered before\cite{qin12,lahtinen17}
in somewhat different contexts.
Investigating the phase diagram of the more general model
$$
H = \sum_{j} -X_j X_{j+1}^\dagger + f Z_j + g Z_{j} Z_{j+1} + g' Z_{j} Z^\dagger_{j+1} + {\rm h.c.}
$$ 
would be very interesting, both in the chiral as well as the non-chiral case\cite{burello14}. 
Finally, it would be interesting to investigate the relation with parafermionic topological
phases, which have attracted quite some attention during the recent years,
see for instance\cite{bondesan,jf14,xu17}.

The interacting transverse field Ising model, Eq.~\eqref{hamil EP} for general $h$ and $U$, is related to the Axial Next-Nearest Neighbour Ising(ANNNI) model, whose phase diagram has been studied thoroughly\cite{selke}. Those studies are related to the large $s$ limit of the PE line and its dual version. The phase diagram of ANNNI model is quite rich and has, for instance, an incommensurate phase. As we showed in Sec.~\ref{numerics}, the large $l$ limit of the $Z_3$ line also has a multi-critical point. In this light, it would be interesting to study the phase diagram of $Z_3$ model and its dual.
       
{\em Acknowledgments ---}
We would like to thank L.~Mazza, C.~Mora, N.~Regnault and D.~Schuricht
for interesting discussions. This work was sponsored, in part, by the Swedish
Research Council.

\appendix

\section{Exact excited states along the PE-line}
\label{app:excited-state}

In this appendix, we prove that the states Eq.~\eqref{eq:e4-states},
$\vert E = 4, \pm \rangle$, are indeed exact
excited states of the Peschel-Emery Hamiltonian for the case with
periodic boundary conditions and an even number of sites, $L = 2N$.
We recall that
\begin{align*}
\vert E = 4, \pm \rangle &= O^-_{\rm tot} \vert E=0;\pm\rangle \\\nonumber
&=4 \mathcal{N}_{\pm} \sum_{j=1}^{N}
\sum_{i_1\cdots i_N = \pm } g_{i_1}\cdots g_{i_{j-1}}e_{-} g_{i_{j+1}} \cdots  g_{i_N} \ .
\end{align*}
where we introduced the notation $O^-_{\rm tot} = \sum_{j=1}^{N} O_j^-$.
We then have that
\begin{align*}
H \vert E = 4, \pm \rangle &=
H  O_{\rm tot}^- \vert E = 0, \pm \rangle \\
&=  O_{\rm tot}^- H \vert E = 0, \pm \rangle 
+ [ H, O_{\rm tot}^- ] \vert E = 0, \pm \rangle \\
&= [ H, O_{\rm tot}^- ] \vert E = 0, \pm \rangle \ .
\end{align*}
It is straightforward to evaluate the commutator
\begin{align*}
[ H, O_{\rm tot}^- ]  = -2i \bigl(
& + \sigma_1^x \sigma^y_2 - \sigma_1^y \sigma^x_2 \\
& - \sigma_2^x \sigma^y_3 + \sigma_2^y \sigma^x_3 \\
& \vdots \\
& - \sigma_L^x \sigma^y_1 + \sigma_L^y \sigma^x_1 \bigr) \ .
\end{align*}
To find the action of the commutator on the ground state, we need to know
\begin{align*}
-2 i (\sigma_1^x \sigma^y_2 - \sigma_1^y \sigma^x_2) \vert \uparrow\uparrow \rangle &= 0 \\
-2 i (\sigma_1^x \sigma^y_2 - \sigma_1^y \sigma^x_2) \vert \downarrow\downarrow \rangle &= 0 \\
-2 i (\sigma_1^x \sigma^y_2 - \sigma_1^y \sigma^x_2) \vert \uparrow\downarrow \rangle &= 
-4 \vert \downarrow\uparrow \rangle \\
-2 i (\sigma_1^x \sigma^y_2 - \sigma_1^y \sigma^x_2) \vert \downarrow\uparrow \rangle &= 
4 \vert \uparrow\downarrow \rangle ,
\end{align*}
resulting in
\begin{align*}
-2 i (\sigma_1^x \sigma^y_2 - \sigma_1^y \sigma^x_2) \vert g_+ \rangle &= 0 \\
-2 i (\sigma_1^x \sigma^y_2 - \sigma_1^y \sigma^x_2) \vert g_- \rangle &= 4 \vert e_- \rangle \ .
\end{align*}
This in turn means that
\begin{align*}
H \vert E = 4, \pm \rangle &=
8 \mathcal{N}_{\pm}(\mathbf{1}-T) \times \\
& \sum_{j=1}^{N}
\sum_{i_1\cdots i_N = \pm } g_{i_1}\cdots g_{i_{j-1}}e_{-} g_{i_{j+1}} \cdots  g_{i_N} 
\end{align*}
where $T$ is the operator that translates the system by one site and we do not sum over
$i_j$. Because
$$
\sum_{j=1}^{N}
\sum_{i_1\cdots i_N = \pm }
g_{i_1}\cdots g_{i_{j-1}}e_{-} g_{i_{j+1}} \cdots  g_{i_N}
$$
is a state with momentum $K=\pi$, as can be verified directly, it follows that
\begin{equation}
H \vert E = 4, \pm \rangle =
4 \vert E = 4, \pm \rangle \ ,
\end{equation}
which we wanted to show. That the other states given in the main text also are exact
excited states can be verified in a similar manner.

\section{Two-site parafermion operator}
\label{app:two-site-parafermion-operator}

In this appendix, we construct the most general parafermion operator that
permutes the three parity ground states Eq.~\eqref{z3paritygs} of the model
Eq.~\eqref{Z3 line}, for arbitrary parameter $r > 0$. We write this operator in the
basis $\{ | 00 \rangle, | 01 \rangle, | 02 \rangle, |10 \rangle, \ldots, |22\rangle\}$.
The operator $O(r)$ we are after should change the sectors as
\begin{align}
\label{lowering-conditions}
O (r)\vert E=0 ; 1 \rangle &= \vert E=0 ; \bar\omega \rangle \nonumber\\
O (r)\vert E=0 ; \omega \rangle &= \vert E=0 ; 1 \rangle \\
O (r)\vert E=0 ; \bar\omega \rangle &= \vert E=0 ; \omega \rangle \ , \nonumber
\end{align}
which is how $X_1$ acts in the case $r = 1$.
This means that $O(r)$ should consist of operators of the form $X_1$,
$X_2$, $X_1^\dagger X_2^\dagger$, $Z_1 X_1$, etc.. In total, there are 27 such operators.
Alternatively, there are 27 non-zero entries in the matrix representation of $O(r)$.
We present the operator in terms of the latter. A convenient labeling turns out to be
\begin{equation}
O(r) = 
\begin{pmatrix}
0 & b_{2;3} & 0 & b_{1;3} & 0 & 0 & 0 & 0 & b_{3;3} \\
0 & 0 & c_{1;2} & 0 & c_{3;2} & 0 & c_{2;2} & 0 & 0 \\
a_{3;1} & 0 & 0 & 0 & 0 & a_{2;1} & 0 & a_{1;1} & 0 \\
0 & 0 & c_{1;1} & 0 & c_{3;1} & 0 & c_{2;1} & 0 & 0 \\
a_{3;3} & 0 & 0 & 0 & 0 & a_{2;3} & 0 & a_{1;3} & 0 \\
0 & b_{2;2} & 0 & b_{1;2} & 0 & 0 & 0 & 0 & b_{3;2} \\
a_{3;2} & 0 & 0 & 0 & 0 & a_{2;2} & 0 & a_{1;2} & 0 \\
0 & b_{2;1} & 0 & b_{1;1} & 0 & 0 & 0 & 0 & b_{3;1} \\
0 & 0 & c_{1;3} & 0 & c_{3;3} & 0 & c_{2;3} & 0 & 0 \\
\end{pmatrix} \ .
\end{equation}
Because it is possible in the $Z_2$ case to write the corresponding operator
using real parameters, we make the same assumption here.
Apart from the conditions Eq.~\eqref{lowering-conditions}, the operator $O$ should
satisfy $O(r)^\dagger = O(r)$ and $O(r)^3 = 1$. The former condition means that the parameters
$a_{i;j}$, $b_{i;j}$ and $c_{i;j}$
form three sets of three orthonormal vectors. So if $\vec{a}_1 = (a_{1;1},a_{1;2},a_{1;3})^T$ etc,
we have $\vec{a}_i^T \cdot \vec{a}_j = \delta_{i,j}$ and similar for the other two sets.
Each of these three sets is constrained by
one of the equations in Eq.~\eqref{lowering-conditions}. In particular, the vectors lie on the intersection
of a sphere and a plane; for each set of orthonormal vectors, there are two such planes. The structure
of the constraints Eq.~\eqref{lowering-conditions} is such that their is a solution. In fact, for each set
of orthonormal vectors, the solution is parametrized by an angle.
Explicitly, these solutions take the form
(using the parameters $c= \sqrt{1+2 r^4}$ and $d=\sqrt{2r^2+r^4}$)
\begin{align*}
a_{1;1} &= \frac{2r^3 + (c d + r^2)\cos(\phi_1) + (-d+c r^2)\sin(\phi_1)}{2 c d} \\
a_{1;2} &= \frac{2r^3 + (-c d + r^2)\cos(\phi_1) + (d+c r^2)\sin(\phi_1)}{2 c d} \\
a_{1;3} &= -\frac{r(-r^3 +\cos(\phi_1) + c \sin(\phi_1) )}{c d} \\
a_{2;1} &= \frac{2r^3 + (-c d + r^2)\cos(\phi_1) - (d+c r^2)\sin(\phi_1)}{2 c d} \\
a_{2;2} &= \frac{2r^3 + (c d + r^2)\cos(\phi_1) + (d-c r^2)\sin(\phi_1)}{2 c d} \\
a_{2;3} &= \frac{r(r^3 -\cos(\phi_1) + c \sin(\phi_1) )}{c d} \\
a_{3;1} &= \frac{r(1-r^3 \cos(\phi_1)+d r \sin(\phi_1))}{cd} \\
a_{3;2} &= -\frac{r(-1+r^3 \cos(\phi_1)+d r \sin(\phi_1))}{cd} \\
a_{3;3} &= \frac{r^2(1+2r \cos(\phi_1))}{cd} \\
%\end{align*}
%\begin{align*}
b_{1;1} &= \frac{2r^3 + (c d + r^2)\cos(\phi_2) + (d - c r^2)\sin(\phi_2)}{2 c d} \\
b_{1;2} &= \frac{2r^3 + (-c d + r^2)\cos(\phi_2) + (d+c r^2)\sin(\phi_2)}{2 c d} \\
b_{1;3} &= -\frac{r(-1 +r^3\cos(\phi_2) + d r \sin(\phi_2) )}{c d} \\
b_{2;1} &= \frac{2r^3 + (-c d + r^2)\cos(\phi_2) - (d+c r^2)\sin(\phi_2)}{2 c d} \\
b_{2;2} &= \frac{2r^3 + (c d + r^2)\cos(\phi_2) + (-d+c r^2)\sin(\phi_2)}{2 c d} \\
b_{2;3} &= \frac{r(1-r^3 \cos(\phi_2) + d r \sin(\phi_2) )}{c d} \\
b_{3;1} &= \frac{r(r^3- \cos(\phi_2)+ c  \sin(\phi_2))}{cd} \\
b_{3;2} &= -\frac{r(-r^3 + \cos(\phi_2)+ c \sin(\phi_2))}{cd} \\
b_{3;3} &= \frac{r^2(1+2r \cos(\phi_2))}{cd} \\
%\end{align*}
%\begin{align*}
c_{1;1} &= \frac{r^2(1+(1+r^2)\cos(\phi_3))}{d^2} \\
c_{1;2} &= \frac{r^2(1-\cos(\phi_3)+d \sin(\phi_3))}{d^2} \\
c_{1;3} &= \frac{r(r^2-r^2\cos(\phi_3)-d \sin(\phi_3))}{d^2} \\
c_{2;1} &= -\frac{r^2(-1+\cos(\phi_3)+d \sin(\phi_3))}{d^2} \\
c_{2;2} &= \frac{r^2(1+(1+r^2)\cos(\phi_3))}{d^2} \\
c_{2;3} &= \frac{r(r^2-r^2\cos(\phi_3)+d \sin(\phi_3))}{d^2} \\
c_{3;1} &= \frac{r(r^2- r^2\cos(\phi_3)+ d  \sin(\phi_3))}{d^2} \\
c_{3;2} &= \frac{r(r^2 -r^2\cos(\phi_3)-d \sin(\phi_3))}{d^2} \\
c_{3;3} &= \frac{r^2(r^2+2 \cos(\phi_3))}{d^2}
\end{align*}

Finally, the condition $O(r)^3 = 1$ leads to the constraint that $\phi_1 + \phi_2 + \phi_3 = 0$. This
leaves a two-parameter family of solutions for the operator $O(r)$. There are three rather special
solutions, namely $\phi_1 = \phi_2 = \phi_3 =-2\pi/3, 0, 2\pi/3$. In the limit $r=1$, when the model
reduces to $h^{Z_3}_{j,j+1} = \bigl[-X_j X^\dagger_{j+1} + {\rm h.c.}\bigr] + 2$,
the operator $O(1)$ becomes
$X_1$, $X^\dagger_1 X^\dagger_2$, $X_2$ in these three cases respectively.
One could hope that the form
of $O(r)$ in the two cases $\phi_1 = \phi_2 = \phi_3 = \pm 2 \pi/3$ would give a hint for the
possible form of two parafermion operators that are exponentially localized at the edges in the
case of longer chains. However, the already rather complicated form of this operator in the two-site
case makes it hard to guess the general form for larger system sizes.

\end{document}